# Investigation of Correction Method of the Spacecraft Low Altitude Ranging

LIU Jing-Lei(刘靖雷) SHEN Chao(沈超) WU Shi-Tong(吴世通) HUANG Wei(黄伟)

Beijing Institute of Space Mechanics & Electricity, Beijing 100094, China

**Abstract:** γ-ray altitude control system is an important equipment for deep space exploration and sample return mission，its main purpose is a low altitude measurement of the spacecraft based on Compton Effect at the moment when it lands on extraterrestrial celestial or sampling returns to the Earth's land, and an ignition altitude correction of the spacecraft retrograde landing rocket at different landing speeds. This paper presents an ignition altitude correction method of the spacecraft at different landing speeds, based on the number of particles γ-ray reflected field gradient graded. Through the establishment of a theoretical model, its algorithm feasibility is proved by a mathematical derivation and verified by an experiment, and also the adaptability of the algorithm under different parameters is described. The method provides a certain value for landing control of the deep space exploration spacecraft landing the planet's surface.

## 1 Introduction
### 1.1 The purposes of spacecraft low altitude measurement and altitude correction

Manned spaceflight and deep-space exploration are two major scientific exploration works carried out by China in the field of aerospace, what they all have in common is the presence of EDL (Entry, Descent, Landing) process, namely the need to return to Earth or the spacecraft landed on the moon. When the spacecraft near the planet's surface time, especially in the moment of contact with the surface of the planet, the relative velocity of the spacecraft and the planet's surface is expected to be zero. In order to achieve this goal, the spacecraft is equipped with a retrograde landing rocket spacecraft for the final step process of EDL[1-5].

Because the spacecraft's retrograde landing rocket has a working time and specific impulse, the spacecraft requires the retrograde landing rocket to start work in a specific altitude to ensure that retrograde landing rocket just ends work at the moment when the spacecraft landing to the ground at a certain speed. When the spacecraft fall speed changes, retrograde landing rocket ignition altitude should correspondingly change to ensure that at other landing speeds, the expectation speed of spacecraft landing on the surface of the planet can be zero[1-4].

Due to the different landing surface of different celestial body is not completeness flat, what the laser altimeter or mechanical touch altimeter measures are all the altitude value for a particular point which can not reflect the real landing point within a certain region. Gamma photons altimeter emits gamma photons diffused through certain area and the receiver receives the statistical value of the reflected photon in the area to reflect accurately the altitude of the place[5-9].

### 1.2 The needs of Spacecraft near Earth altimeter and altitude correction

In general, the spacecraft uses the parachutes or the engine to decelerate during EDL process. The bottom of the spacecraft is equipped with retrograde landing rocket, when the spacecraft fell to the ground from the altitude $H_0$ and the rate of decline in the return capsule is for $v_0$, the γ photon altimeter below the spacecraft issues a directive and the retrograde landing rocket starts ignition work and reduces the rate of decline in the return capsule to $v_1$ within the altitude range of brake stroke Ha. Before landing the retrograde landing rocket should first be turned off to avoid the "explosion" phenomenon caused by blocking the vents. To simplify the calculation, assuming that the thrust $F_r$ of the retrograde landing rocket keeps constant within the altitude of the entire brake stroke $H_a$, and during this period the resistance of the parachute system is not considered, then the ignition altitude $H_0$ of the retrograde landing rocket is determined by the following three formulas.

$$H_S = \frac{1}{2g}\left(v_2^2 - v_1^2\right) \quad (1)$$

$$F_r H_a = \frac{1}{2}m\left(v_0^2 - v_1^2\right) + mgH_a \quad (2)$$

$$H_0 = H_a + H_S \quad (3)$$

In the above formula, $m$ is the quality of the return capsule. $H_s$ stands for the remaining altitude of braking return capsule and $v_2$ for the return capsule landing speed. In particular, due to the impact of landing field environment, altitude and the atmosphere during the returns process[3], the spacecraft's landing speed is not a changeless value, but changes within a range[1]. In order to adapt to the above style, the spacecraft ignition altitude correction is required, i.e., at high landing speeds, the ignition altitude is higher while at lower landing speeds, the ignition altitude is lower. Automatic compensation of the ignition signal altitude, Altitude is determined by the following formula:

$$H = H_0 + K(V - V_H) \quad (4)$$

In the above formula, $H$ is the altitude of the ignition. $H_0$ is the desired altitude. $V_H$ is the desired speed. $K$ is correction coefficient. $V$ is the actual rate of decline. Generally, $K$ value is between 0.15~0.27m/m/s.

## 2 The spacecraft low altitude ranging method based on Compton effect

### 2.1 The physical model

As shown in Fig 1, the transmitter transmits a dose of gamma rays to the ground. Diffuse through the surface of the planet, the gamma photons into light pulses through the NaI crystal, the photo multiplier tube converts the optical signal into a weak current pulse signals[3]. Furthermore, amplified converted current signals into pulse signals.

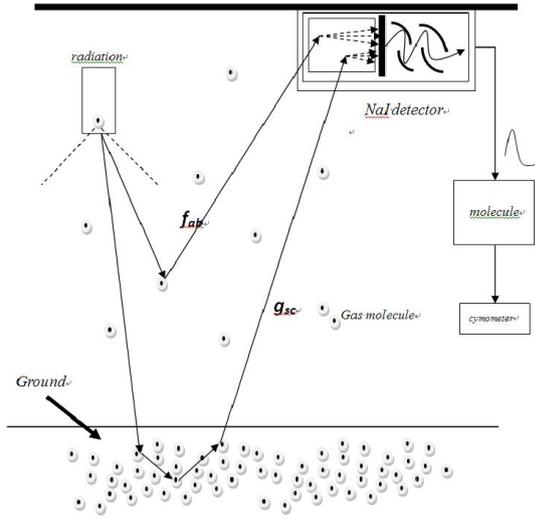

Fig.1. The physical process of altitude measurement

Fig 2 shows the gamma altimeter altitude measurement mathematical model. At the $H_0$ altitude transmitter continuous emission to the ground gamma rays by $2\psi$ angle[3]. At the altitude $H=H_0+B$, receiving terrestrial gamma photons back scattered by the receiving processor. $d$ is the transmitter and the horizontal distance between the crystals.

Then captured by the NaI crystal gamma photons back scattered from the ground density can be expressed as:

$$\varphi_P = \frac{I_0 \cdot K_0 \cdot V}{4\pi} \cdot \int_{\theta=0}^{\Psi} \int_{\beta=0}^{2\pi} \frac{\varepsilon_\gamma(E_0,\rho,Z,\theta_0,\theta_1,\beta) \cdot \sin\theta}{[d^2 + H^2 \cdot tg^2\theta + (H_0+B)^2 - 2d \cdot H_0 \cdot tg\theta \cdot \cos\beta]} d\theta \cdot d\beta \quad (5)$$

In the above formula: $I_0$ for activity radioactive sources, $K_0$ is the decay constant $3.7 \times 10^{10}$ times / sec, $V$ is the decay coefficient, $\varepsilon_\gamma(E_0,\rho,Z,\theta_0\theta_1,\varphi)$ as inverse scattering probability, $E_0$ is photon radiation energy sources, and is the role of the effective atomic number of the material, $\beta$ is azimuth angle; $\theta_0\theta_1$ is the angle of incidence and the inverse scattering angle, $H_0$ is the ground distance to the radiation source.

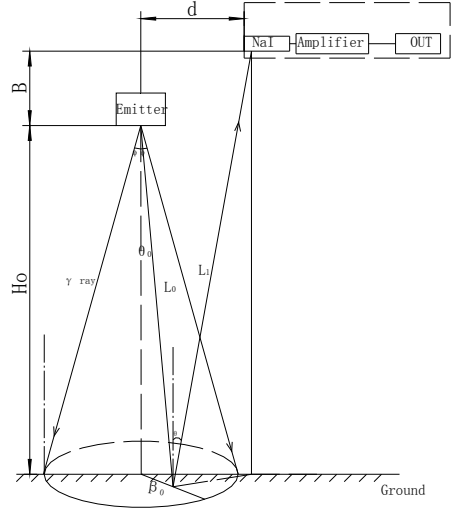

Fig.2. The physical model of altitude measurement

Particularly in $2\psi$, $B$, $d$ and other parameters are known conditions, NaI effective area of the transistor unchanged, the relationship between the number of photons($Np$) received by NaI crystal and the flux density can be simplified to:

$$N_p = \varphi_p \cdot S \quad (6)$$

Effective count by NaI crystal conversion a frequency is expressed as of:

$$f(h) = N_p \cdot A(E,h) \quad (7)$$

In the above formula $A(E,h)$ is a function of frequency counts that the receiver convert it at different altitudes when the gamma photons incident NaI crystal. By the above formula can be obtained that at different altitudes, the receiver output different frequencies. We can calculate the altitude of the spacecraft which depending on the frequency.

### 2.2 Experimental Verification

We can draw that the receivers at different altitudes received different gamma photon number from the previous analysis. And we can calculate the altitude of the spacecraft which depends on the frequency. Therefore, the test system is designed in this paper. Using $Cs^{137}$ as a reference source, according to the relevant literature, gamma energies between 30~200keV with an altitude information[3]. The receiver receives the signal of the energy spectrum of gamma particles segment. We install the transmitter and receiver in the experimental model to research on the theory model. The out sole structures were winched to 20m, 10m, 7m, 5m, 3.9m, 3.3m, 2.6m, 2.2m, 1.9m, 1.6m, 1.3m, 1.1m, 0.7m, 0.5m altitude (the vertical distance to the ground test center). Altitude error caused by experimental equipment is: (5mm+0.2%H). The pulse signal receiver connected by a cable to ground frequency meter. Measurement strobe frequency is set to 1$s$. All experiments used as a radiation source. The static H-V curve as shown below, which describes the change between the different receivers .

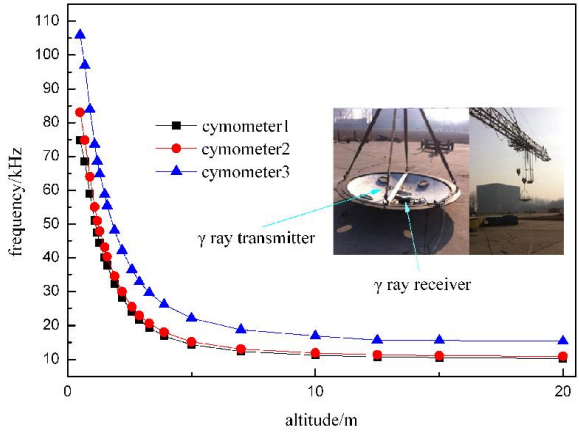

Fig.3. The static H-V curve

Through the above data can be seen that the spacecraft receive different frequency at different altitudes, it can be judgment altitude by frequency. The F-V curve has a good rate of frequency change in the range of 0.5~5m. and the rate of frequency change is not very obvious when the altitude preponderate over 5m. It can also be seen that different receiver measures different frequency and has a collective migration tendency.

## 3 Altitude correction model
### 3.1 Physical Model

H-F function is represented by f(h). Suppose time series TS which has a length of a fixed value T is at time coordinate system memory space. TS does uniform motion in the time coordinate system. That is to say, the moment is longer and longer at both ends of the TS, but the TS has fixed length. Suppose TS has a short space length T0 at both ends and the TS start time and end time are remembered as 0 and T, denoted as [0, T], TS, then start time and end time at both ends of the space are respectively [0, T0] and [T-T0, T].

Suppose F(h,v) is a function of the integral of f(h), then F(h,v) is expressed as:

$$F(h,v) = \int_{h}^{h+T_0 V} f(h)dh - \int_{h+TV-T_0 V}^{h+TV} f(h)dh \quad (8)$$

Assuming the ignition is desired altitude $h_0$ when the speed is $v_0$, there is:

$$F(h_0,v_0) = \int_{h_0}^{h_0+T_0 V_0} f(h)dh - \int_{h_0+TV_0-T_0 V_0}^{h_0+TV_0} f(h)dh = C_0 \quad (9)$$

Assuming the desired altitude value $h_0$ is constant while the rate of decline is different, F($h_0$,v) expression is:

$$F(h_0,v) = \int_{h_0}^{h_0+T_0 V} f(h)dh - \int_{h_0+TV-T_0 V}^{h_0+TV} f(h)dh = C_V \quad (10)$$

When V> $V_0$, we can draw that $C_V$> $C_0$.

Therefore, it is possible to obtain altitude correction according to the function F(h,v) changes.

### 3.2 Experimental Verification

Test system is established to validate the physical model. The test system measures ignition altitude with different rate of decline speed. The experiment selects the 6.6m/s (expectation ignition altitude is 0.73M), 7.4m/s (expected ignition altitude 0.89m) test value of speed. Determination of T values is 100ms, T0 is selected as 5ms and 10ms. Four trials were conducted at different speeds to verify whether the ignition altitude has correction function. The experimental data are shown as follows:

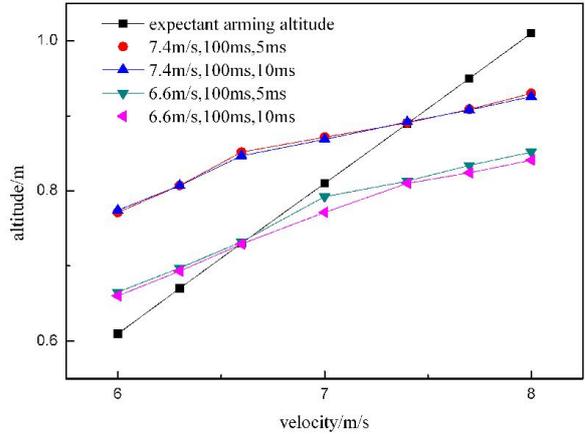

Fig.4. Altitude correction of two specific speed

According to the above experiment, K=0.2. We calibrated the speed of 8.0m/s, 7.7m/s, 7.4m/s, 7.0m/s, 6.6m/s, 6.3m/s, 6.0m/s and then carry on the experiment. The data obtained are listed in the following table:

Table1. The experiment data

| Velocity /m/s | Expectant arming altitude /m | Testing arming altitude /m |
|---|---|---|
| 10.0 | 1.53 | 1.523 |
| 9.7 | 1.47 | 1.463 |
| 9.4 | 1.40 | 1.391 |
| 9.0 | 1.32 | 1.299 |

| 8.6 | 1.24 | 1.226 |
| 8.3 | 1.18 | 1.208 |
| 8.0 | 1.12 | 1.118 |
| 7.7 | 1.02 | 1.003 |
| 7.4 | 0.96 | 0.947 |
| 7.0 | 0.87 | 0.877 |
| 6.6 | 0.79 | 0.795 |
| 6.3 | 0.73 | 0.742 |
| 6.0 | 0.66 | 0.672 |

Ignition altitude of the curve shown in Fig5.

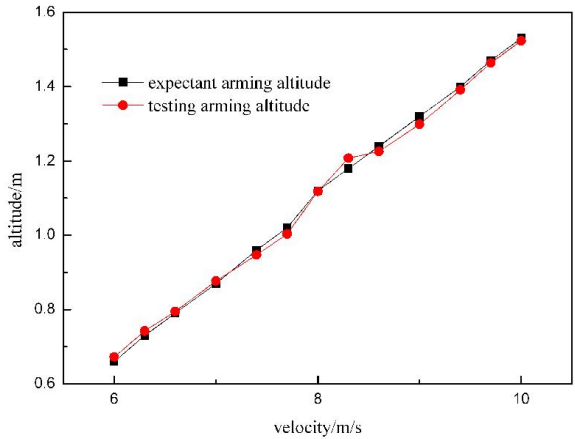

Fig.5. Altitude correction results

## 4 The analysis of H-F curve
### 4.1 Physical Model

As mentioned earlier, γ photon frequency is:

$$N = f\left(\varphi_{pH_0}, S, d, B, \varepsilon, \tau\right) \quad (11)$$

At the same altitude, $\varphi_{pH0}$, S, d, B, ε, τ are unchanged.

The following conclusions can be drawn:

$$N \propto (I_0, E_0, V_{ref}) \quad (12)$$

Where $V_{ref}$ is the reference value circuit. Assume that in the vicinity of known HF curve, the function of the variable changes, such as emission source activity, circuit noise, the temperature drift and so on. We can consider these changes occurring in a continuous smooth curve. When a small change occurs, it can be described as:

$$N = C \cdot N_0 \quad (13)$$

Also can be expressed as:

$$f_n(h) = f_0(h) + Cf_0(h) \quad (14)$$

### 4.2 Data Analysis

The above experimental data are divided into three groups, respectively denoted as $f_1$, $f_2$, $f_3$. And calculate the average of three groups data, It denoted as $f_0$, We calculate the relative differences of three groups data and the average frequency. It can be seen that there groups data are standard differences with the average of three groups data.

In the fig 6, due to the three-way receiver device independent, the difference of different curve is caused by factors but not the gamma source. It can seen in the basic formula 13.

We can get the standard curve by using the formula of HF curve analysis. For different values of C, we can obtain the HF curve set. When product debugging, downloading the parameters according to the parameters of HF curve of the most close to the actual selection of correction parameters can speed up the debugging time.

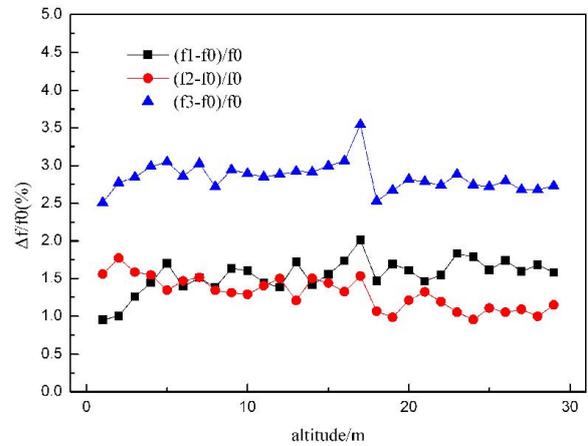

Fig.6. The H-F curves of different relative offset

## 5 Conclusions

In this paper, we discusse the spacecraft altitude measurement and correction method based on the Compton effect, and carrie on a related experiment to verify our research. Experimental data show that the effective altitude and the altitude correction rang is 0.5~1.5m. This method can meet the speed range 6m/s~10m/s. The accuracy is ± 0.05m. It can meet certain engineering need.

E-mail: telleo@163.com